\documentclass[twocolumn]{aastex62}
\graphicspath{{./}{figures/}}
\usepackage{xcolor}

\shorttitle{Vela pulses at mm wavelengths with ALMA}
\shortauthors{Liu et al.}

\begin{document}
\author[0000-0002-2953-7376]{Kuo Liu}
\affiliation{Max-Planck-Institut f\"ur Radioastronomie, Auf dem H\"ugel 69, D-53121 Bonn, Germany}

\author[0000-0003-0000-2682]{Andr\'e Young}
\affiliation{Department of Astrophysics, Institute for Mathematics, Astrophysics and Particle Physics (IMAPP), Radboud University, P.O. Box 9010, 6500 GL Nijmegen, The Netherlands}

\author[0000-0002-7416-5209]{Robert Wharton}
\affiliation{Max-Planck-Institut f\"ur Radioastronomie, Auf dem H\"ugel 69, D-53121 Bonn, Germany}

\author[0000-0002-9030-642X]{Lindy Blackburn}
\affiliation{Center for Astrophysics $\vert$ Harvard \& Smithsonian, 60 Garden Street, Cambridge, MA 02138, USA}
\affiliation{Black Hole Initiative at Harvard University, 20 Garden Street, Cambridge, MA 02138, USA}

\author{Roger Cappallo}
\affiliation{Massachusetts Institute of Technology Haystack Observatory, 99 Millstone Road, Westford, MA 01886, USA)}

\author[0000-0002-2878-1502]{Shami Chatterjee}
\affiliation{Cornell Center for Astrophysics and Planetary Science, Cornell University, Ithaca, NY 14853, USA}

\author[0000-0002-6156-5617]{James M. Cordes}
\affiliation{Cornell Center for Astrophysics and Planetary Science, Cornell University, Ithaca, NY 14853, USA}

\author[0000-0002-2079-3189]{Geoffrey B. Crew}
\affiliation{Massachusetts Institute of Technology Haystack Observatory, 99 Millstone Road, Westford, MA 01886, USA}

\author[0000-0003-3922-4055]{Gregory Desvignes}
\affiliation{Max-Planck-Institut f\"ur Radioastronomie, Auf dem H\"ugel 69, D-53121 Bonn, Germany}
\affiliation{LESIA, Observatoire de Paris, Universit\'e PSL, CNRS, Sorbonne Universit\'e, Universit\'e de Paris, 5 place Jules Janssen, 92195 Meudon, France}

\author[0000-0002-9031-0904]{Sheperd S. Doeleman}
\affiliation{Center for Astrophysics $\vert$ Harvard \& Smithsonian, 60 Garden Street, Cambridge, MA 02138, USA}
\affiliation{Black Hole Initiative at Harvard University, 20 Garden Street, Cambridge, MA 02138, USA}

\author[0000-0001-6196-4135]{Ralph P. Eatough}
\affiliation{Max-Planck-Institut f\"ur Radioastronomie, Auf dem H\"ugel 69, D-53121 Bonn, Germany}

\author[0000-0002-2526-6724]{Heino Falcke}
\affiliation{Department of Astrophysics, Institute for Mathematics, Astrophysics and Particle Physics (IMAPP), Radboud University, P.O. Box 9010, 6500 GL Nijmegen, The Netherlands}

\author{Ciriaco Goddi}
\affiliation{Department of Astrophysics, Institute for Mathematics, Astrophysics and Particle Physics (IMAPP), Radboud University, P.O. Box 9010, 6500 GL Nijmegen, The Netherlands}
\affiliation{Leiden Observatory - Allegro, Leiden University, P.O. Box 9513, 2300 RA Leiden, The Netherlands}

\author[0000-0002-4120-3029]{Michael D. Johnson}
\affiliation{Center for Astrophysics $\vert$ Harvard \& Smithsonian, 60 Garden Street, Cambridge, MA 02138, USA}
\affiliation{Black Hole Initiative at Harvard University, 20 Garden Street, Cambridge, MA 02138, USA}

\author[0000-0002-7122-4963]{Simon Johnston}
\affiliation{CSIRO Astronomy and Space Science, Australia Telescope National Facility, PO Box 76, Epping, NSW 1710, Australia}

\author[0000-0002-5307-2919]{Ramesh Karuppusamy}
\affiliation{Max-Planck-Institut f\"ur Radioastronomie, Auf dem H\"ugel 69, D-53121 Bonn, Germany}

\author[0000-0002-4175-2271]{Michael Kramer}
\affiliation{Max-Planck-Institut f\"ur Radioastronomie, Auf dem H\"ugel 69, D-53121 Bonn, Germany}

\author[0000-0002-3728-8082]{Lynn D. Matthews}
\affiliation{Massachusetts Institute of Technology Haystack Observatory, 99 Millstone Road, Westford, MA 01886, USA}

\author[0000-0001-5799-9714]{Scott M. Ransom}
\affiliation{National Radio Astronomy Observatory, Charlottesville, VA 22903, USA}

\author{Luciano Rezzolla}
\affiliation{Institut f{\"u}r Theoretische Physik, Goethe-Universit{\"a}t Frankfurt, Max-von-Laue-Stra{\ss}e 1, D-60438 Frankfurt am Main, Germany}

\author{Helge Rottmann}
\affiliation{Max-Planck-Institut f\"ur Radioastronomie, Auf dem H\"ugel 69, D-53121 Bonn, Germany}

\author[0000-0002-6514-553X]{Remo P.J. Tilanus}
\affiliation{Leiden Observatory - Allegro, Leiden University, P.O. Box 9513, 2300 RA Leiden, The Netherlands}
\affiliation{Department of Astrophysics, Institute for Mathematics, Astrophysics and Particle Physics (IMAPP), Radboud University, P.O. Box 9010, 6500 GL Nijmegen, The Netherlands}
\affiliation{Netherlands Organisation for Scientific Research (NWO), Postbus 93138, 2509 AC Den Haag , The Netherlands}

\author[0000-0001-8700-6058]{Pablo Torne}
\affiliation{Instituto de Radioastronom\'ia Milim\'etrica, IRAM, Avenida Divina Pastora 7, Local 20, 18012, Granada, Spain}
\affiliation{Max-Planck-Institut f\"ur Radioastronomie, Auf dem H\"ugel 69, D-53121 Bonn, Germany}

\title{Detection of pulses from the Vela pulsar at millimeter wavelengths with phased ALMA}

\correspondingauthor{K.~Liu}
\email{kliu@mpifr-bonn.mpg.de}

\begin{abstract}
We report on the first detection of pulsed radio emission from a radio pulsar
with the ALMA telescope. The detection was made in the Band-3 frequency range ($85-101$~GHz)
using ALMA in the phased-array mode developed for VLBI observations.
A software pipeline has been implemented to enable a regular pulsar
observing mode in the future. We describe the pipeline and demonstrate
the capability of ALMA to perform pulsar timing and searching. We also measure the flux density and polarization properties of the
Vela pulsar (PSR J0835$-$4510) at mm-wavelengths, providing the first polarimetric study
of any ordinary pulsar at frequencies above 32~GHz.
Finally, we discuss the lessons learned from the Vela observations
for future pulsar studies with ALMA, particularly for searches near the
supermassive black hole in the Galactic Center, and the potential of using
pulsars for polarization calibration of ALMA.
\end{abstract}


\section{Introduction} \label{sec:intro}

Pulsars are steep-spectrum radio sources \citep[e.g.,][]{lk04}. As a result, the vast majority of the pulsar population have been discovered at frequencies below 2\,GHz.
Correspondingly, the study of the radio emission has been limited to similar
frequencies, although successful studies have been conducted up to much higher
frequencies. Before 1990, the highest frequency used for successful pulsar
studies was 25\,GHz, while in the 1990s observations were pushed to 32\,GHz
\citep{wjkg93}, 43\,GHz \citep{kjdw97}, and finally 87\,GHz \citep{mkt+97}.
Emission from normal pulsars was later observed at 138\,GHz \citep{tor16}, and
finally, detections of the radio-emitting magnetar PSR J1745$-$2900 were
achieved at frequencies as high as 291\,GHz \citep[for a review, see][]{tor2018}.

Observations at high frequencies will provide a better understanding
of pulsar emission physics and allow for more effective pulsar searches
in highly turbulent environments like the Galactic Center \citep{cl97,lk04,sle+14,ddb+17}.
Previous pulsar studies have suggested that the coherent emission seen at lower
radio frequencies may undergo changes at high radio frequencies \citep{kxj+96}.
This may be understood as a break-down of the coherent radiation mechanism, which can be
expected to occur when the observed wavelength becomes comparable to the
coherence length. The break-down would correspond to a transition from the coherent radio
emission to the incoherent infrared or optical emission \citep{lk04}. At the
same time, a standard model of pulsar emission physics interprets observed
pulse characteristics as the result of a ``radius-to-frequency mapping'', where
higher radio frequencies are emitted from lower emission heights \citep{cor78}.
In the context of this model, performing observations at higher radio frequencies is
equivalent to approaching the polar cap region of the pulsar.

All previous studies of pulsars above 30 GHz have been conducted with
Northern hemisphere radio telescopes, especially the 100-m Effelsberg radio telescope near Bonn,
Germany and the 30-m telescope of the Institut de Radioastronomie
Millim\'etrique (IRAM) on Pico Veleta, Spain (see \citealt{ljk+08} and
\citealt{tde+17} and references therein). With the advent of the Atacama
Large Millimeter/submillimeter Array (ALMA), a large collecting area is
now available to study Southern hemisphere pulsars. Here we report on the
establishment of a fast time-domain capability (hereafter a
``pulsar observing mode'') for ALMA's phased-array system \citep{mcd+18}.
This new pulsar observing mode can be used for observations of compact
objects in the Galactic Center and elsewhere in the Galaxy. We demonstrate
the capabilities of this system with observations of the Vela pulsar.

As one of the brightest radio pulsars in the sky, Vela was one of the
first pulsars discovered \citep{lvm68} despite its relatively fast spin period
($P_{\rm spin} = 89~{\rm ms}$).
Recently, an imaging detection of Vela was made at millimeter wavelengths
using ALMA in its standard interferometry observing mode \citep{mpr+17}. As
shown below, we can now use {\em phased} ALMA (i.e., ALMA in the phased-array mode) to obtain a complementary
study of Vela's pulsed and polarised emission at frequencies of 90\,GHz
and above.

ALMA can also play a key role in probing neutron star populations in the
Galactic Center and using detected objects for the study of spacetime around
the central black hole, the Sgr~A* \citep{lwk+12,pwk16,le17}. Despite previous efforts, no
pulsar in a sufficiently close orbit to Sgr~A* has yet been detected
\citep{wcc+12}. However, the discovery of a rare radio-emitting magnetar
(PSR J1745$-$2900) with a projected distance of only $\approx\! 0.1$~pc from
the Sgr~A* \citep{efk+13} suggests the existence of an
intrinsic pulsar population in the immediate vicinity of Sgr~A*.  PSR
J1745$-$2900 has been studied up to frequencies of 291~GHz \citep{tde+17},
indeed implying that ALMA will be an ideal instrument to study its properties.

The development project described here (the {\em ALMA Pulsar Mode Project},
hereafter APMP) involves a series of steps needed to  acquire phased ALMA data for pulsar studies:
definition and implementation of the appropriate phasing mode, providing the
signal path from the ALMA Phasing Project (APP) system to Mark 6 baseband
recorders, offline resampling and reformatting of data into PSRFITS format
\citep{hsm04}, and development of the backend pulsar/transient processing customized to
ALMA science contexts. The project leveraged software development for pulsar
phased-array modes for the Very Large Array (VLA)\footnote{\url{https://science.nrao.edu/facilities/vla/docs/manuals/oss/performance/pulsar}}
and developments for the Event Horizon Telescope \citep{dwr+08,eht+19}, Black Hole
Camera \citep{gfk+17} projects, and the ALMA Phasing Project \citep{mcd+18}.

\section{Observations}~\label{sec:obs}

Feasibility studies conducted for the APMP used test data obtained in
conjunction with ALMA Phasing Project commissioning runs in 2016 April and 2017 January.
The former provided data used to test the integrity of the Mark 6
to PSRFITS transformation while the latter provided data on the Vela pulsar
(PSR J0835$-$4510) for validation purposes. Detection of the pulsar at a
significance consistent with the sensitivity (A/T) and bandwidth used was the
primary goal in order to demonstrate the feasibility of pulsar and transient
observations with ALMA. Demonstrating the ability to use pulsar observations
for system tests and instrumental polarisation calibration were secondary
goals. Finally, comparing the properties of the Vela pulse profile with those obtained at lower
frequencies allows us to study pulsar emission physics.

\begin{figure}[htbp]
\epsscale{1.2}
\plotone{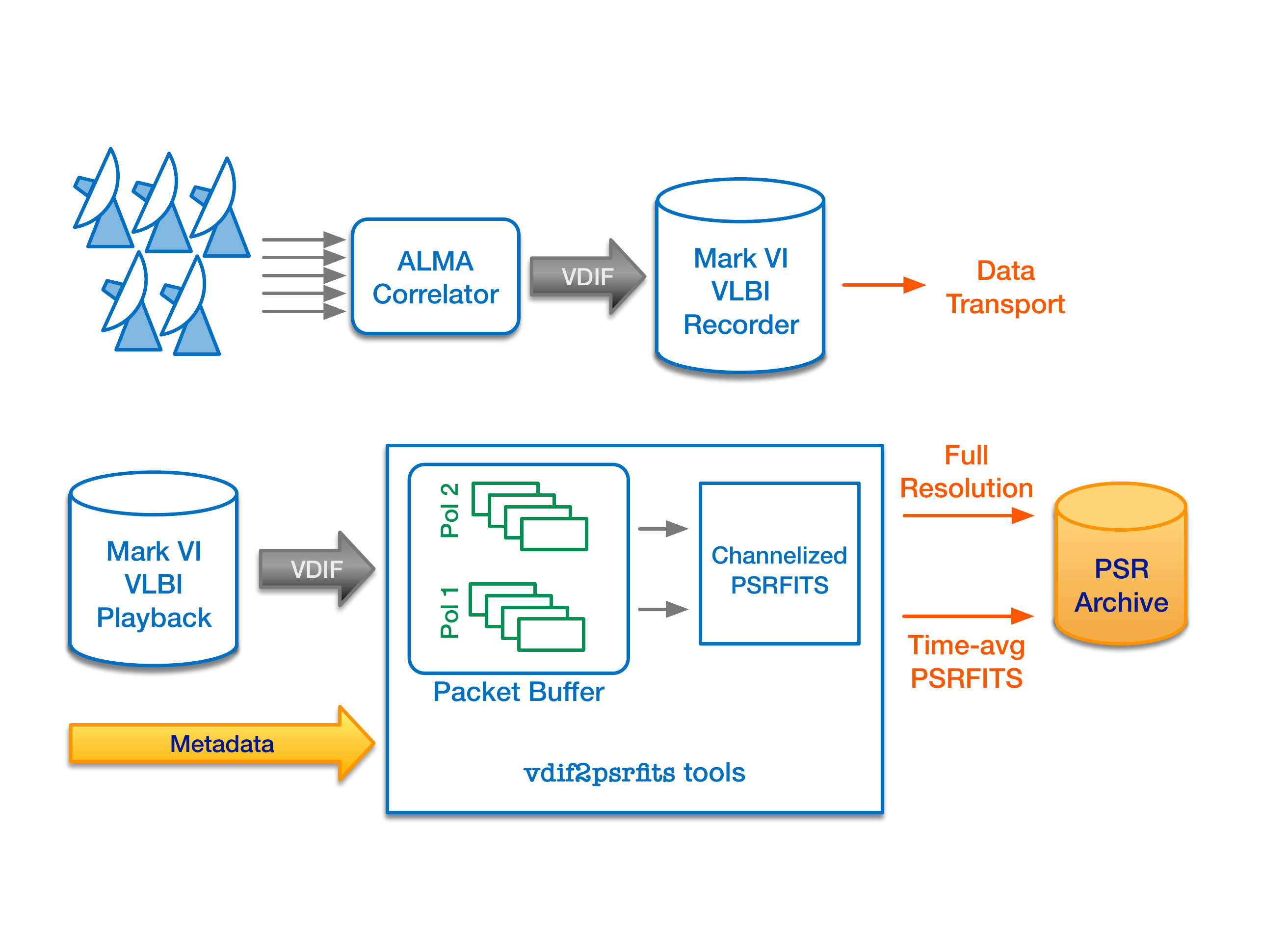}
\caption{\footnotesize
The developed ALMA system for offline processing to produce pulsar/transients
data. The phased ALMA voltage data are recorded on Mark6 VLBI recorders
\citep{wbc+13}. In our study the data were played back and processed at
Cornell University, the Massachusetts Institute of Technology (MIT)
Haystack Observatory, the Center for Astrophysics $|$ Harvard \& Smithsonian (CfA) and the Max Plank Institut f\"ur Radioastronomie (MPIfR) to produce PSRFITS format pulsar data
(\url{https://www.atnf.csiro.au/research/pulsar/psrfits_definition/Psrfits.html}). The box represents the software suites that convert VDIF format data into PSRFITS
data, and are publicly available on
\url{http://hosting.astro.cornell.edu/research/almapsr/}. As described there, two pipelines of such have been developed independently at MPIfR and CfA. \label{fig:dataflow}}
\end{figure}

The observation of the Vela pulsar was conducted with ALMA using Band-3 on 2017 January 29 under excellent weather conditions. The observation spanned approximately 40~min.
In total, 37 12-m antennas were phased up to form a tied-array beam, delivering
a collecting area equivalent to a 73-m parabolic dish.
The entire observation was divided into 8 individual scans where the telescope
was alternately pointed at Vela and a bright neighboring phase calibrator
(J0828$-$3731, 0.87\,Jy measured on 2015-12-25 and 8\,deg separation).
``Active phasing'' mode was deployed during the calibrator scans, where the phasing solution was updated every 18.192\,s. During the scans on the Vela
pulsar, the latest phasing solution from the last calibrator scan was adopted
and kept unchanged (the so-called ``passive phasing'' mode). During the
observation, the baseband data streams from two sidebands (each consisting of
two sub-bands) were recorded, providing 4$\times$2-GHz sub-bands centered at
86.268, 88.268, 98.268 and 100.268\,GHz, respectively. Each sub-band was
subdivided into $32\times 62.5$-MHz frequency channels. The data were then
processed off-line to yield intensity detections for all Stokes parameters with
a time resolution of 8\,$\mu$s and packed in \textsc{PSRFITS} format in search
mode (channelized time series). For the purpose of detecting the Vela pulsar,
the data were folded to form 10-s sub-integrations, using an ephemeris obtained from low frequency observations around the same period of time.
Details on the data flow and the pre-processing can be found in
Figure~\ref{fig:dataflow}.

\section{Results}
\begin{figure*}[htbp]
\centering
\includegraphics[scale=0.9]{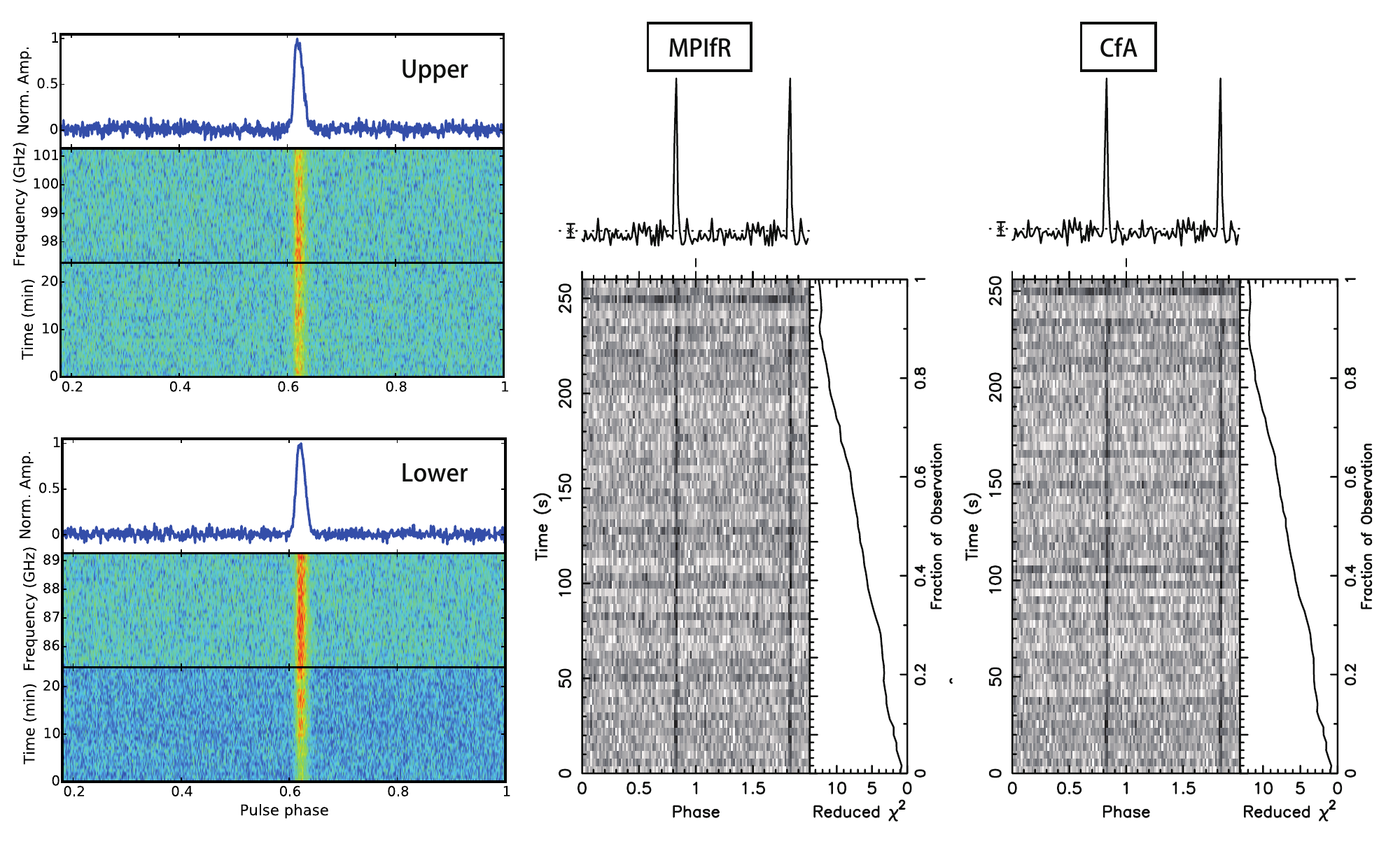}
\caption{\footnotesize
Left panel: Pulse profiles detected in the lower (bottom, $85.268$-$89.268$\,GHz) and upper
sideband (top, $96.268$-$101.268$\,GHz) of ALMA Band-3, using the MPIfR pipeline. Middle and right panel: Comparison of pulse profile achieved from the MPIfR and CfA pipeline \citep[folded using the \textsc{presto} software package,][]{rem02}. In short, the MPIfR pipeline makes power detection of all four Stokes parameters in frequency domain, while the CfA pipeline derives total intensity power directly from the state counts. With the same section of data, the detection from these two pipelines shows highly consistent measures of detection significance in total intensity and the shape of the pulse profile of the Vela pulsar. The product from the MPIfR pipeline is used for the data analysis in the rest of this paper. \label{fig:prof-lo-hi}}
\end{figure*}

The Vela pulsar has been successfully detected after folding the search mode data. Figure~\ref{fig:prof-lo-hi} shows the integrated pulse profiles
obtained from the lower and upper sideband, respectively. An effective
integration time of 25\,min yields a peak signal-to-noise ratio of the detection of
$\sim$50 for the lower and $\sim$40 for the upper sideband. The signal was
detected at frequencies up to 101.268\,GHz, the highest radio frequency that
the pulse profile of Vela has ever been seen. Overall, the ALMA detection of the
Vela pulsar provides a successful demonstration
of the APMP and provides the basis for future execution of our
primary scientific motivation: searches and follow-up observations of pulsars
and transient sources in the Galactic center.

\subsection{Timing}
\label{ssec:timing}
\begin{figure}[htbp]
\epsscale{1.2}
\plotone{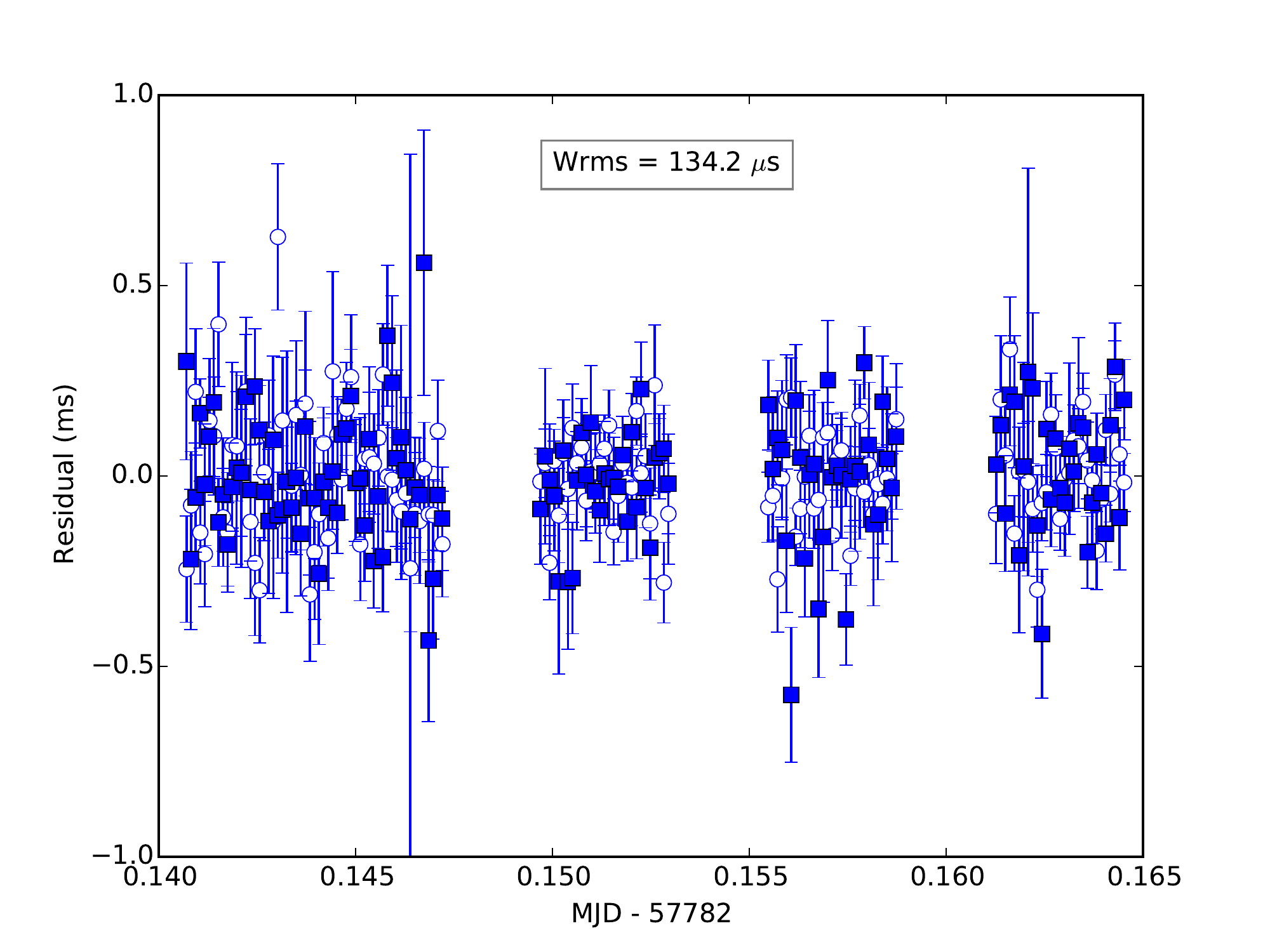}
\caption{\footnotesize
         Timing residuals of Vela pulse profiles derived from 10-s sub-integrations. The open circles and squares represent residuals from the lower and upper sideband, respectively.
         \label{fig:res}}
\end{figure}

The utility of pulsars as precision clocks, the so-called pulsar timing
technique, requires high timing stability during the process of data
recording. To examine such a capability of the phased ALMA data, we carried
out a timing analysis with the 10-s sub-integrations from both sidebands. For
each of them, we averaged the pulse profile over frequency, and calculated its
time-of-arrival (TOA) with the canonical template-matching method \citep{tay92}.
Figure~\ref{fig:res} shows the timing residuals obtained by subtracting
the predictions of the ephemeris (obtained independently from low-frequency
observations) from the TOAs. No time offset was seen
between the four individual scans. A weighted root-mean-square (rms)
timing uncertainty of 134\,$\mu$s has been achieved, which is in general
consistent with the TOA errors expected from radiometer noise. This shows
that a timing precision of order $100$\,$\mu$s that was used in the simulation
of \cite{lwk+12}, is in fact possible for pulsar observations with ALMA at
3-mm wavelengths.

\subsection{Search capability} \label{ssec:search}

\begin{figure}[htbp]
\epsscale{1.2}
\plotone{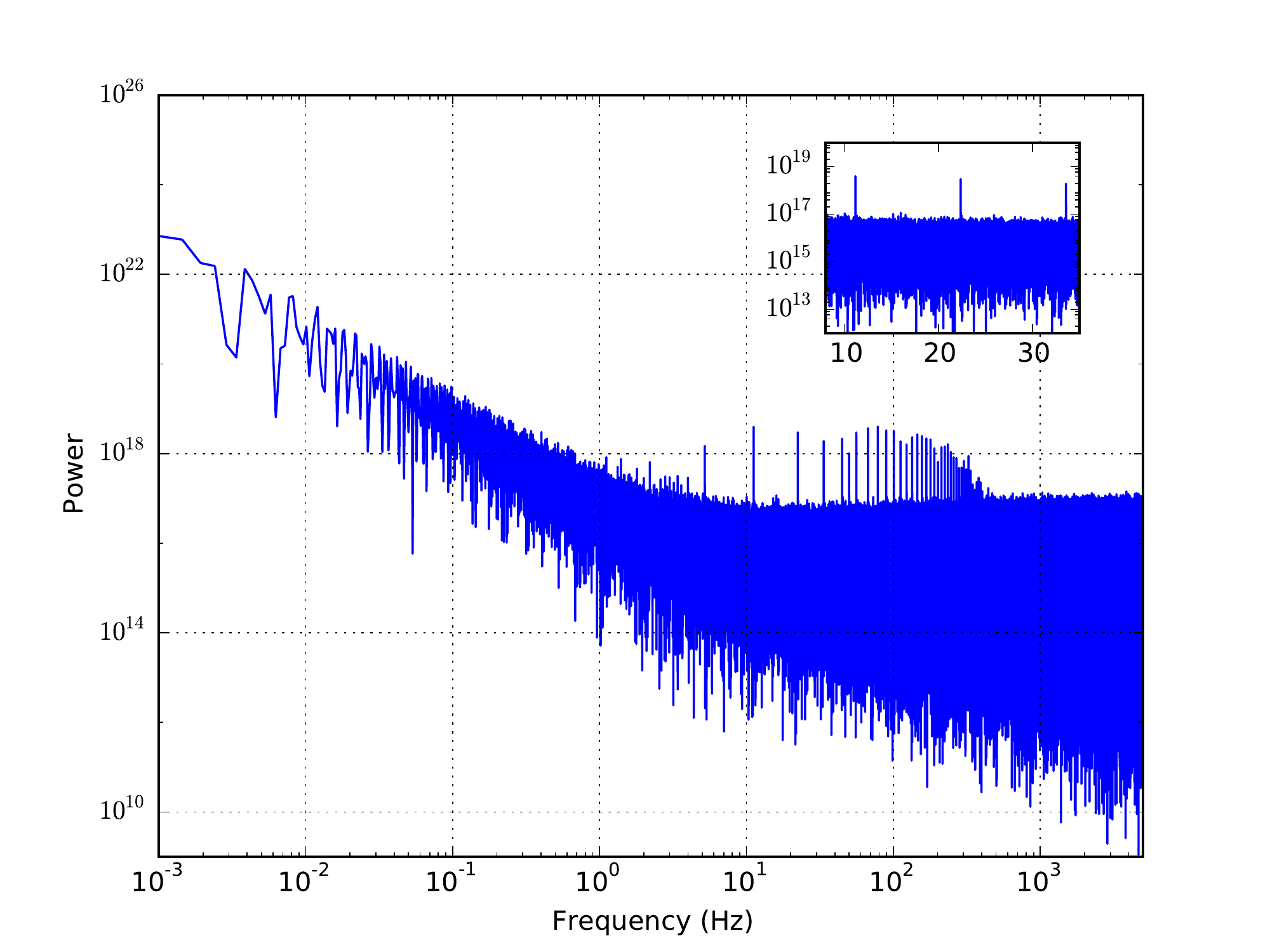}
\caption{\footnotesize
Fourier power spectrum of the overall time series formed by combining data from
all four individual Vela scans. For each scan, the time series was first averaged in
frequency. The inset shows a zoomed-in region of the power spectrum for frequencies
between 5-35\,Hz, with linear scale on the x-axis. Note that the spinning frequency of the Vela pulsar is approximately 11.2\,Hz.
\label{fig:pspec}}
\end{figure}

To demonstrate the capability of the APMP for searching for periodic signals in
the data, we used the overall dataset to directly carry out a blind search for the Vela pulsar. We first
averaged the time series from all four individual sub-bands for each scan, and
combined the time series from all four scans, with the power mean padded in the gaps among the scans. Then
a periodicity search was performed using the
\textsc{presto}\footnote{\url{https://www.cv.nrao.edu/~sransom/presto/}}
software package \citep{rem02}. Figure~\ref{fig:pspec} shows the power
spectrum of the combined time series. The low frequency noise, mainly caused by fluctuations of the power level in the time series,
starts to become significant for frequencies below 5\,Hz.
Meanwhile, the overall power level of the rest of the spectrum is mostly flat.
The fundamental spinning frequency of the Vela pulsar (around 11.2\,Hz) and its higher order harmonics
are clearly seen in the Figure~\ref{fig:pspec} inset. The periodicity search
resulted in 28 candidates with detection significance above $3\sigma$.  The
fundamental spin frequency of the Vela pulsar was the top candidate and most
of the rest were higher harmonics. At frequencies of 50 and 1~Hz, periodic
signals were detected and associated with the electricity power cycle and
cryogenic pump's cycle, respectively. Another signal was detected at 2.6~Hz, and
its origin is unknown but is likely to be terrestrial, as it was also seen in
scans of the calibrator\footnote{The ALMA baseline correlator is clocked
(precisely) at 125 MHz with microprocessor interrupts at 16\,ms and timing
signals at 48\,ms---so signals commensurate with these are likely to have been
produced in the correlator.}.

\subsection{Pulsar properties}
\label{ssec:prof}
We can use the obtained data to study the properties of the Vela pulsar at frequencies between 80 and 100~GHz and compare those to Vela's properties at
lower frequencies, and to those of PSR~B0355+54 (to be presented in future work), the only other normal pulsar
detected at similar frequencies \citep{mkt+97,tor16}. This will allow us to gauge the prospects of future pulsar observations with ALMA and further our understanding of pulsar emission physics.

\subsubsection{Flux Density}
\label{ssec:flux}

The recorded data of the phase calibrator J0828$-$3731 allow us to estimate the
mean flux density of the Vela pulsar profile. For each
individual scan we produced imaging detection of the phase calibrator and
measured its flux density at the lower and upper sideband, respectively, which
gave consistent values with those from the ALMA calibrator catalog\footnote{\url{https://almascience.nrao.edu/sc/}}. Then for each individual Vela scan, we used the closest calibrator scan to calibrate the flux density of the pulsed emission, by using the standard flux calibration formula \citep{lk04}. This gave us a mean flux density of $0.99\pm0.17$\,mJy and
$0.69\pm0.12$\,mJy at 87.268 and 99.268\,GHz, respectively, where the error bars
represent the actual standard deviation of all four individual measurements. Using the
ordinary ALMA interferometry data recorded in parallel during the observation
and calibrated following the dedicated procedures developed for phased ALMA
\citep{Goddi2019}, we also produced image detections of the Vela pulsar from all
individual scans, which delivered mean flux density measurements of
$0.82\pm0.09$\,mJy and $0.67\pm0.05$\,mJy at 87.268 and 99.268\,GHz,
respectively. These are fully consistent with the measurements derived from the
phased ALMA data.

\subsubsection{Profile evolution}
\label{ssec:profevolv}
\begin{figure}[htbp]
\includegraphics[scale=0.4,trim={0 5cm 0 3cm},clip]{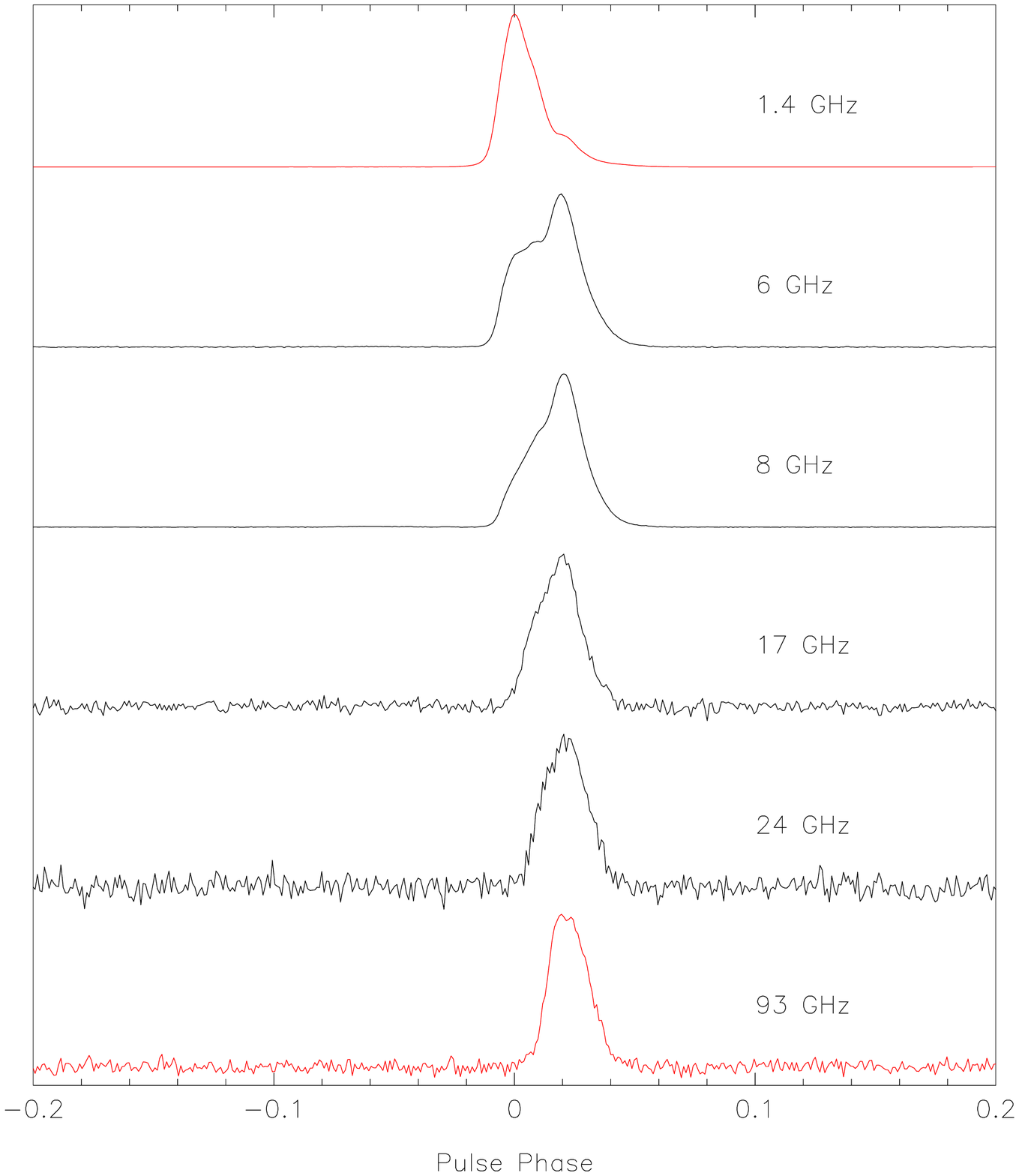}
\caption{\footnotesize
Vela profiles measured as a function of frequency. The bottom profile (red) is
the result of the addition of the two ALMA sidebands. It is aligned in time,
using the data's timestamps, with the 1.4 GHz profile shown on top (red), which
resulted from an observation with the Parkes Radio Telescope at 1.4 GHz made on 2017 January 8. Note
that using nearly contemporaneous observations with Parkes, we minimize any
confusing time delay potentially caused by rotational instabilities known as
``timing noise.'' Indeed, the apparent delay of 1.8~ms of the ALMA profile
with respect of the pulse peak seen at 1.4~GHz can be explained by a distinct
profile evolution. This becomes clear when adding the profiles (black) observed
and aligned by \cite{kjlb11}. See text for details.
\label{fig:profile}}
\end{figure}

%
The lack of a drift during the folding of the data (see
Fig.~\ref{fig:prof-lo-hi}) and the apparent timing stability (Section
\ref{ssec:timing}), imply that the data timestamps are reliable. Going further,
we can compare the phase of the pulse arrival time at 1.4\,GHz (obtained at the Parkes Radio Telescope), defined by that
of the main pulse peak (identified with phase zero), with that of the pulse
peak seen with ALMA at mm-wavelengths. Figure~\ref{fig:profile} demonstrates that the latter is delayed by about 1.8~ms with respect to the main pulse peak at 1.4~GHz. This offset is explained by comparing the profiles to those at intermediate
frequencies. We use those presented by \cite{kjlb11}, following their alignment
based on a separation of the profile into individual components\footnote{\citet{kjlb11} modelled the profiles at different frequencies
as a sum of a number of symmetric components represented by scaled von Mises
functions.  They found that a set of the same four component fits to all
frequencies whilst keeping the width and separation of each component fixed.
We aligned their solution relative to our ALMA and Parkes observations by eye.}.
As can be seen from Figure~\ref{fig:profile} the main component seen at
1.4\,GHz (and below) becomes progressively weaker at high frequencies and is
undetectable at ALMA frequencies. The component remaining is the second,
weaker component of the 1.4\,GHz profile. This is consistent with the
identification of the dominant 1.4-GHz component seen in the top profile of
Figure~\ref{fig:profile}, with a so-called ``core component'' \citep[see e.g.][]{jvkb01} which tends to have steeper spectra.
Indeed, \cite{kjlb11} measure a spectral index of $-2.7\pm0.1$ for this
component, compared to $-1.5\pm0.2$ for the component seen with ALMA\footnote{Note that we identify \cite{kjlb11}'s component C and/or D with
the ALMA component, for which a spectral index of $-1.56\pm0.06$ and
$-1.5\pm0.2$ was determined, respectively. We adopt $-1.5$ with its larger
uncertainty here.}. Extrapolating from the 24\,GHz flux density
\citep[3.4\,mJy, ][]{kjlb11}, we therefore expect a flux density of
$0.94\pm0.14$\,mJy at 87\,GHz and $0.82\pm0.31$\,mJy at 99\,GHz. This is in
perfect agreement with the flux density measurement we obtained (see Section~\ref{ssec:flux}).

\subsubsection{Polarization properties}
\label{ssec:pol}
\begin{figure}[htbp]
\plotone{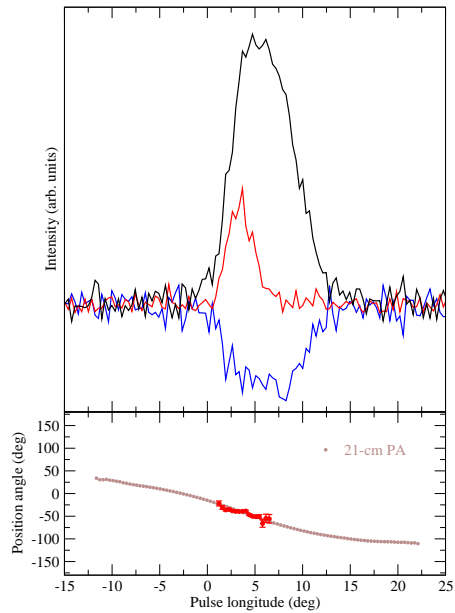}
\caption{\footnotesize
Polarisation properties of the Vela pulsar as observed with ALMA, when adding
the Stokes vectors measured for the lower and upper sideband. The top panel
shows the total power (black), the linearly polarised emission component (red),
and the circularly polarised emission (blue). The lower panel shows the
position angle of the linearly polarised component as a function of those pulse
phases, where the linear intensity exceeds $1.5\sigma$ of the off-pulse region.
We also indicated the position angle swing measured at 1.4\,GHz, corrected for
Faraday rotation to infinite frequency.
\label{fig:pol}}
\end{figure}

The MPIfR pipeline allows for the extraction of full-Stokes information
from ALMA baseband data. We find that Vela still shows some significant linear and
circular polarisation. Assuming that the observed polarisation characteristics
are similar to those observed at 24\,GHz, we can use the data by \cite{kjlb11}
as a polarisation template and perform a system calibration without needing to
make any assumptions on ``ideal feeds'' or additional constraints on degeneracy
in the system parameters\footnote{Here we chose only to fit for differential gain and phase, as the leakage in ALMA Band-3 was shown to be no more than a few percentage \citep{Goddi2019}.} \citep{van06,sbj+17}. The result is shown in Figure~\ref{fig:pol}, where we show an expanded region around the pulse. Here, the degree of linear polarisation ($\sim 20\%$) is lower than at 24~GHz but the position angle is perfectly consistent with the 1.4~GHz data (after correcting for Faraday rotation). This change in degree of polarisation, while maintaining a constant PA swing
(believed to be tied to the magnetic field and the viewing geometry of the pulsar), is perfectly consistent with overall trends in other pulsars \citep{lk04}. The sign of circular polarisation is identical to that at lower frequencies and its fraction relative to total intensity is similar to at 24~GHz.

\section{Discussion} \label{sec:dis}
We have demonstrated the capability of phased ALMA for the study of pulsars. The use
of phased ALMA in combination with a passive phasing mode will enable future
pulsar searches, as well as timing, polarisation, and emission studies.
This opens up new science possibilities, especially in the southern hemisphere,
which is completely unexplored for pulsars at frequencies above 30 GHz.

Given the typical steep spectrum of pulsars, they will generally be
far too weak at ALMA frequencies to allow active phasing of the ALMA
array on the pulsar itself. The periodicity search experiment carried out in this paper used a
passive phasing mode. Our results demonstrate the high data quality that can
be achieved using this mode of phased ALMA for pulsar studies. Meanwhile, in
the data collected in active phasing scans on a bright calibrator source,
we noticed significant systematics in the time series which are associated with
the phasing cycles. This issue will be investigated further in a forthcoming
paper.  That means that it will be highly preferable for future pulsar
observations with phased ALMA to be conducted in passive phasing mode,
irrespective of the flux density of the source.

As shown in our analysis, it is possible to use a pulsar as a calibrator to
understand and calibrate the polarisation of phased ALMA data. In particular,
since the Vela pulsar exhibits apparent circular polarisation at
3-mm wavelengths, it could potentially help to better estimate the leakages in a
linear feed system. Additionally, the low rotation measure of this pulsar \cite[${\rm RM} = 31.4~{\rm rad~m}^{-2}$,][]{jhv+05}, guarantees that the contamination by Faraday rotation during calibration process is negligible at 3-mm wavelengths. To carry out this experiment, a long track of the Vela pulsar needs
to be conducted in order to cover a wide range of parallactic angle. Then a calibration can
be performed by following the approach described in \cite{van04,van06}, after
which a polarisation template of the Vela pulsar at the given frequency will be
constructed. For calibration afterwards, one would only need a short scan on
the Vela pulsar and match it to the template.

\acknowledgments
We thank A.~Evans, T.~Remijan and F.~Stoehr for the help to stage our data on the ALMA science portal\footnote{The data are available via: \url{https://almascience.eso.org/alma-data/enhanced-data-products/vela-pulsar-j0835-4510}.}. KL, RW, GD, RPE, HF, CG, MK, LR, PT acknowledge the financial support by the European Research Council for the ERC Synergy Grant BlackHoleCam under contract no. 610058. SC and JMC acknowledge support from the National Science Foundation (AAG 1815242). Work on this project at the Smithsonian Astrophysical Observatory was funded by NSF Grant AST-1440254. This paper makes use of the following ALMA data: ADS/JAO.ALMA\#2011.0.00004.E. ALMA is a partnership of ESO (representing its member states), NSF (USA) and
NINS (Japan), together with NRC (Canada), MOST and ASIAA (Taiwan), and KASI
(Republic of Korea), in cooperation with the Republic of Chile. The Joint ALMA
Observatory is operated by ESO, AUI/NRAO and NAOJ.
The National Radio Astronomy Observatory is a facility of the National Science
Foundation operated under cooperative agreement by Associated Universities, Inc.
The ALMA Phasing Project was principally supported by a Major Research Instrumentation award from the National Science Foundation (award 1126433) and an ALMA North American Development Augmentation award to Cornell University; the ALMA Pulsar Mode Project was supported by an ALMA North American Study award.

\facility{ALMA}

\software{PRESTO}

\bibliographystyle{aasjournal}
\bibliography{almapsr}

\end{document}